\pdfoutput=1
\documentclass[runningheads]{llncs}
\usepackage{graphicx}
\begin{document}
\title{TestLab: An Intelligent Automated Software Testing Framework}
%
%\titlerunning{Abbreviated paper title}
% If the paper title is too long for the running head, you can set
% an abbreviated paper title here
%
\author{Tiago Dias\orcidID{0000-0002-1693-7872} \and
Arthur Batista \orcidID{0009-0006-2354-0773}
Eva Maia\orcidID{0000-0002-8075-531X} \and
Isabel Praça\orcidID{0000-0002-2519-9859}}
\authorrunning{T. Dias et al.}
% First names are abbreviated in the running head.
% If there are more than two authors, 'et al.' is used.
%
\institute{Research Group on Intelligent Engineering and Computing for Advanced Innovation and Development (GECAD), Porto School of Engineering (ISEP), 4200-072 Porto, Portugal \\
\email{\{tiada,1171006,egm,icp\}@isep.ipp.pt}}
\maketitle              % typeset the header of the contribution
\begin{abstract}
The prevalence of software systems has become an integral part of modern-day living. Software usage has increased significantly, leading to its growth in both size and complexity. Consequently, software development is becoming a more time-consuming process. In an attempt to accelerate the development cycle, the testing phase is often neglected, leading to the deployment of flawed systems that can have significant implications on the users daily activities. This work presents TestLab, an intelligent automated software testing framework that attempts to gather a set of testing methods and automate them using Artificial Intelligence to allow continuous testing of software systems at multiple levels from different scopes, ranging from developers to end-users. The tool consists of three modules, each serving a distinct purpose. The first two modules aim to identify vulnerabilities from different perspectives, while the third module enhances traditional automated software testing by automatically generating test cases through source code analysis.

\end{abstract}

\section{Introduction}
\label{sec:intro}

Software systems have become a crucial asset from a professional and personal standpoint. What was previously considered superfluous has now become much more ingrained to a point where it has become a dependency. Consequently, avid users have become compelled to trust these systems, which can be dangerous if they are faulty. The trustiness of software directly relates to its quality~\cite{Torkar2006}. Higher software quality is typically associated with more reliable and secure systems, which in turn boosts users' confidence in interacting with them. The quality and effectiveness of software testing directly impacts the security of the software application. Incomplete or insufficient testing can leave vulnerabilities undetected, compromising the overall security of the system. Therefore, it is crucial to ensure the quality of software used in real-world tasks. One way to achieve this is through software testing, which plays a vital role in the software development cycle by enhancing the quality of the software.

Software testing focuses on evaluating software to determine its correctness and ensure it meets the project requirements. The main objective of software testing is to identify defects within the software, that when correctly addressed may improve the software's quality, reliability, and longevity~\cite{Torkar2006}. It consists of executing the Software Under Test (SUT) and monitoring it for errors, bugs, and other issues, and is usually performed by developers or testers. Despite the availability of automation and management tools, software testing is often skipped or rushed due to its time-consuming nature and the perceived increase in development costs. However, the long-term cost associated with the lack of testing and subsequent issues can also be significant. 

According to recent reports, the total cost of poor software quality in the United States (US) was around 2 trillion dollars~\cite{Krasner2020}. Additionally, the National Institute of Standards and Technology (NIST) found that an average bug detected in early development stages takes approximately five hours to be fixed, whereas, in a post-product release, it takes around 15.3 hours. Therefore, in addition to improving the quality of the Software, testing decreases development time and costs. Regarding the use of automated testing tools, 44\% of Information Technology companies automated 50\% of testing in the year of 2020, with 24\% seeing an increase in Return Of Investment~\cite{TrueList2022}. Additionally, automated tools have proven to be helpful in uncovering vulnerabilties often missed by testers because of their randomness~\cite{Torkar2006}. Despite the integration of automated software testing tools as part of the software testing process, during the first quarter of 2022, over 8000 vulnerabilities were discovered and documented~\cite{Comparitech2022}. 

The concerning number of software defects recently reported, underscore the need for improved software testing processes. As such, this work introduces the TestLab framework, an automated software testing tool designed to address multiple levels of software testing from different scopes ranging from developers to end-users. The framework aims to ensure the delivery of high-quality software systems throughout the software development cycle. By covering testing from different perspectives, TestLab enhances the efficiency and effectiveness of the software testing process, ultimately improving the overall quality of software systems.

This paper is organized as follows. Section~\ref{sec:SoTA} reviews the fundamentals of software testing and related work presented in recent literature. Section~\ref{sec:proposal} describes the conceptualization of the proposed framework. Finally, Section~\ref{sec:conclusion} summarises and concludes the work, describing new research lines to be explored in the future.
\section{State of the Art}
\label{sec:SoTA}

Software testing is the verification and validation process of software, whose goal is to answer two respective questions: (i) is the right system being built?, and (ii) is the system being rightfully built?~\cite{Sommerville2010}. The validation question is answered by the product owner and end-users that validate the user requirements and the satisfaction of the developed solution, respectively~\cite{Torkar2006}. The verification question can be answered by ensuring, to a certain extent, that the software was correctly built and tested. 

\begin{figure}
    \centering
    \includegraphics{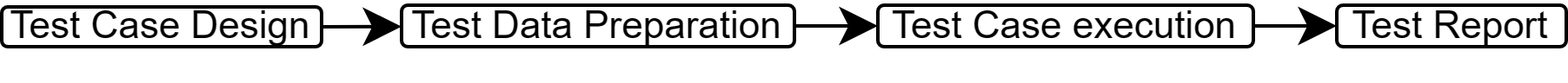}
    \caption{Software Testing Process Model}
    \label{fig:soft_proc}
\end{figure}

In this sense, software testing plays a critical role in the software development process, as it usually appears as the phase that evaluates the correctness, completeness and accuracy of the developed work, revealing its defects. Nonetheless, compliance with best engineering practices is essential for ensuring the testability of software under development. Testability, which refers to the ease of testing a system, is a critical software quality attribute. Neglecting testability, along with insufficient testing, can lead to the failure of major projects. Therefore, prioritizing testability and conducting comprehensive testing are crucial for delivering high-quality software systems~\cite{Uddin2019}. Errors and defects raised by software testing should not be disregarded nor considered an adversity towards the software being developed~\cite{Uddin2019}. Software testing involves the creation of test cases, which are executed using test data to generate testing reports, as depicted in  Fig.~\ref{fig:soft_proc}. However, this process can be arduous and time-intensive, particularly for large and intricate systems under test. Hence, automation is necessary to efficiently ensure software quality while minimizing the resources required~\cite{Dudekula2012}.

Automated Software Testing (AST) can be applied in many different perspectives and at various levels, with the goal of testing in different ways multiple parts of the system. AST is often divided in three different types: (i) black-box, (ii) white-box, and (iii) grey-box. The former is used to evaluate the SUT from a end user's perspective by providing only information regarding the interaction with the system, without revealing its inner-workings~\cite{Murnane2001,Nidhra2012}. White-box testing evaluates the SUT from the developers point of view by examining the internal structure and logic of a software program~\cite{Nidhra2012,Saglietti2008,Martin-Lopez2021}. Grey-box testing combines the former types described, leveraging from their techniques to design more effective and efficient test cases~\cite{Rajamanickam2019}. Moreover, AST can be applied in multiple testing levels ranging from fine-grained testing, where the system is tested in units, to more general testing simulating the actions of a user. Testing a system in multiple levels is relevant, as the lack of it can jeopardise the quality of the software being developed~\cite{Garousi2013}. The levels of testing are the following~\cite{Uddin2019}:
\begin{enumerate}
    \item \textbf{Unit Testing} – The system is broken down into units, and these are tested independently.
    \item \textbf{Integration Testing} – The integration between units is tested to ensure it is defect-free.
    \item \textbf{System Testing} – The system is tested in its entirety, focusing on multiple testing aspects inherent to publicly accessible systems.
    \item \textbf{Acceptance Testing} – The last level of testing, which is performed by the end-users, reflects the acceptability of the system.
\end{enumerate}

Testing a system in multiple levels ensures the correctness of the system at different granularity levels. Though, combined with multiple testing types, the produced test cases can simulate testing performed from different perspectives, which will ultimately provide different insights regarding the software's quality. 

\subsection{Related Work}

The ever-growing interest, size and complexity of software systems have increased the testing effort. As a result, this process is many times skipped, in order to fasten the software development cycle. However, the release of faulty software can also produce unpleasant outcomes, creating an urge to invest in AST. Nowadays, most used and well-known AST tools rely on scripts to automate the execution of test scripts and the generation of reports to deliver faster and less prone to error testing. Additionally, they may even be integrated into Continuous Integration/Continuous Delivery pipelines to ensure the quality of the software under development throughout its development. Nonetheless, these tools still require human effort to write the test cases into the scripts and to function as oracle~\cite{Sommerville2010}. In this sense, the integration of Artificial Intelligence (AI) in the software testing process is promising, as it may be capable of fulfilling the human tasks that are found in the current state of AST. 

Table~\ref{tab:ast_tools} presents a comprehensive comparison of the most common AST tools based on the findings of multiple recent reviews~\cite{Gamido2019,Hanna2018,Umar2019,Singh2014,Kakaraparthy2017}, describing the license, testing levels considered and if they leverage AI.
\begin{table}[]
\centering
\caption{Common automated software testing tools.}\label{tab:ast_tools}
\begin{tabular}{|p{3.3cm}|p{1.3cm}|p{5.8cm}|p{1.4cm}|}
\hline
Tool and Frameworks        & License     & Testing Levels                        & AI-Aided \\ \hline
Selenium                   & Open Source & System Testing and Acceptance Testing & Yes         \\ \hline
Unified Functional Testing & Licensed    & System Testing                        & Yes         \\ \hline
Test Complete              & Licensed    & System Testing and Acceptance Testing & Yes         \\ \hline
Ranorex                    & Licensed    & System Testing and Acceptance Testing & Yes         \\ \hline
Watir                      & Open Source & System Testing                        & No          \\ \hline
Load Runner                & Licensed    & System Testing                        & Yes         \\ \hline
\end{tabular}
\end{table}
Even though these tools are capable of automating software testing, they still require some degree of human expertise during the test case generation, as their use of AI is merely for assistance rather than automation. Moreover, most of them focus on high-level testing of the SUT, performing only at the system testing and acceptance testing levels. 

Regarding the integration of AI in the AST process to automate the test case generation, multiple works have been developed and reviewed~\cite{Vinicius2019,Vitorino2023,Battina2019}. Vitorino et al.~\cite{Vitorino2023} present the challenges associated with the automatic test case generation faced in different testing type conditions. Their systematic review concludes that in a white-box setting most approaches resort to metaheuristics to constraint the input generation in order to achieve different test cases. Whilst in a black-box or grey-box setting, since the lack of data is the root problem, most approaches rely on the information regarding the SUT provided by the testers.
Vinicius et al.~\cite{Vinicius2019} reviews multiple works that achieve successful results using Machine Learning (ML) to further automate software testing, focusing on test case generation, refinement and evaluation, with some works mentioning the necessity for explainable algorithms in this context.

In conclusion, the findings show that even though there has been recent development towards software testing automation resorting to more intelligent methods, no prior work has addressed the development of a comprehensive framework comprising multiple testing methods that operate at various levels and in different settings (e.g. white-box, grey-box, and black-box). Such a framework would be capable of performing fully automated continuous testing of a SUT while integrating and analyzing the results from different testing methods.
\section{Proposed Framework}
\label{sec:proposal}

TestLab is an automated and intelligent software testing framework that comprises multiple testing methods at different levels and various perspectives, such as white-box, black-box, and grey-box. This paper presents TestLab and its automated testing methods. 

\begin{figure}
    \centering
    \includegraphics{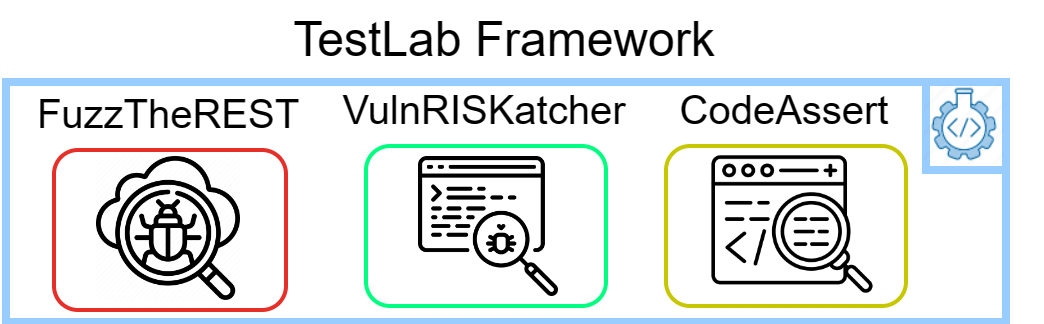}
    \caption{TestLab architecture.}
    \label{fig:testlab}
\end{figure}
TestLab consists of 
three modules, as depicted in Fig.~\ref{fig:testlab}: FuzzTheREST, VulnRISKatcher, and CodeAssert. The first two aim to automate the process of analyzing the SUT for potential vulnerabilities in order to verify the software's security.
FuzzTheREST is a fuzzer that focuses on testing Representational State Transfer (REST) Application Programming Interfaces (APIs) from a black-box point of view, where only data unrelated to the system's inner workings are utilized by the testing tool. VulnRISKatcher is a ML-based tool that has the capability to test source code developed in various programming languages in order to identify vulnerabilities. Lastly, CodeAssert is the part of the framework that attempts to automate the test scripts writing process by analyzing the source code in a white-box fashion.
The subsequent subsections provide a more detailed description of each of these modules.

\subsection{FuzzTheREST}

As previously described in Section~\ref{sec:SoTA}, software testing is composed of multiple methods capable of evaluating the quality of different software characteristics. Fuzzy testing is an automated software testing technique that consists in generating random malformed input and testing it against the SUT to find vulnerabilities and code defects~\cite{Shen2021}. This technique is usually applied in a black-box or grey-box setting leading to different input generation techniques. Typically, black-box fuzzers are quite na\"{i}ve, as their testing input is randomly generated. This makes the vulnerability discovery an indefinitely time consuming task, rendering this kind of fuzzer unfeasible considering the size and complexity of modern day software systems~\cite{Godefroid2007}. On the other hand, grey-box fuzzers leverage existing information regarding the SUT's inner workings. Even though these fuzzers are usually capable of computing faulty input faster, they require more in-depth knowledge about the SUT.

FuzzTheREST is a fuzzer that performs intelligent fuzzy testing on REST APIs in a black-box fashion, with the goal of finding software defects. The fuzzer proposed relies on Reinforcement Learning (RL) to solve the search problem associated with discovering the right input values and combinations to uncover vulnerabilities in the software. The goal of the tool is to find vulnerabilities in web APIs in a platform-independent way, hence the need for the black-box approach. To tackle the na\"{i}veness associated with black-box fuzzers, the tool is equipped with a RL algorithm capable of driving the input that is generated relying on multiple mutation methods and using the API's feedback as guidance. Additionally, the tool can be fine-tuned to include grey-box information by integrating code execution analysis tools to obtain real-time execution metrics, such as coverage.

\begin{figure}
    \centering
    \includegraphics{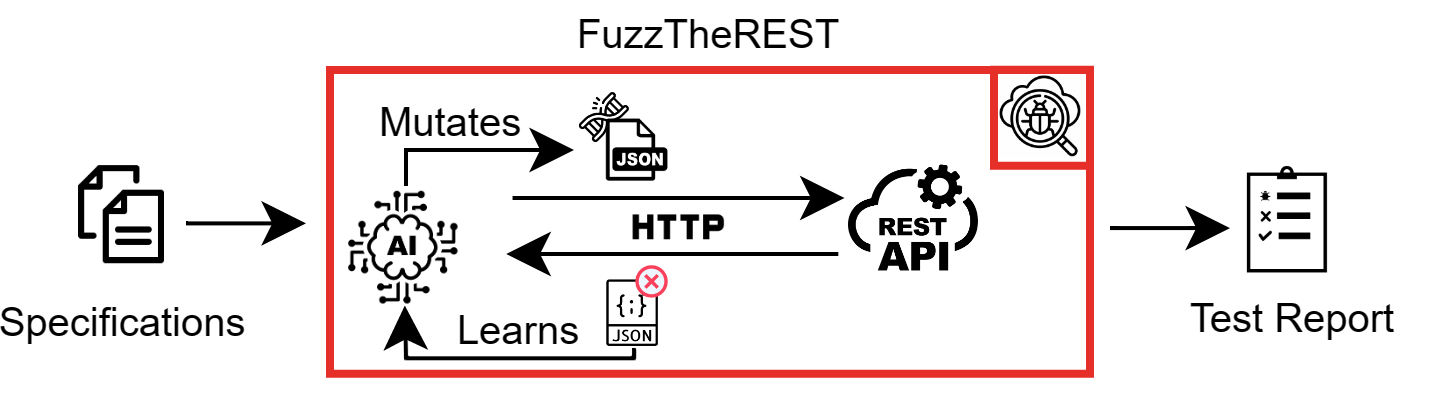}
    \caption{FuzzTheREST pipeline.}
    \label{fig:FTR_pipe}
\end{figure}

Fig.~\ref{fig:FTR_pipe} describes the execution flow of the system. The tester provides all necessary information regarding the SUT, which includes the OpenAPI specification file, the scenarios file and the parameters of the RL algorithm. The OpenAPI specification file containing information to establish interaction with the target API, and the scenarios file describing user-defined combinations between functionalities to find vulnerabilities that may result of state changes. Afterwards, the tool starts fuzzying the API under test, iterating through each user-defined scenario. At the beginning of each scenario, a RL algorithm is instantiated with the chosen parameters and the initial population composed of the necessary input values is randomly generated. The fuzzer then starts sending the requests to the API to evaluate the generated input, using its feedback to mutate it accordingly. Throughout the fuzzy testing process, the algorithm records vulnerabilities, generated input values, and respective mutations to balance exploration and exploitation. This balance ensures that the search for faulty input values does not get stuck in a suboptimal solution (\textit{local optimum}) allowing for a more efficient exploration of the solution space. After testing all scenarios, the fuzzer exports a test report file containing multiple metrics and the vulnerabilities found.

\subsection{VulnRISKatcher} 

% Each year, more vulnerabilities are discovered that can compromise services. Detecting and fixing bugs during the test phase will reduce the time and money spent in the future. 
Every year, the number of vulnerabilities found contaminating software systems is rising, consequence of the ever-growing digital transformation. Their early detection and identification is crucial to ensure quality of the software and to significantly reduce the software development and maintenance costs.

Currently, code review tools mainly rely on static approaches and are typically introduced at the system test level. However, these tools are often less effective and may require a comprehensive understanding of the code context, resulting in a complex and time-consuming process \cite{Clang}. As mentioned in Section~\ref{sec:SoTA}, the potential of ML in software testing is significant \cite{Nazim2022}, as it can overcome the limitations of static code review tools. By analyzing software metrics such as code complexity and code coverage, ML algorithms can learn patterns in the code that may suggest potential vulnerabilities, thereby improving the accuracy and efficiency of identifying them.

VulnRISKatcher is designed to accurately identify and classify vulnerabilities in source code and their associated risks. This tool uses ML techniques for code verification, which do not require the full context of the code, making it applicable at different levels of testing. VulnRISKatcher utilizes multiple ML techniques and leverages lexical code from diverse programming languages to train the models. This approach enables the solution to identify and classify various types of vulnerabilities accurately.

\begin{figure}
    \centering
    \includegraphics{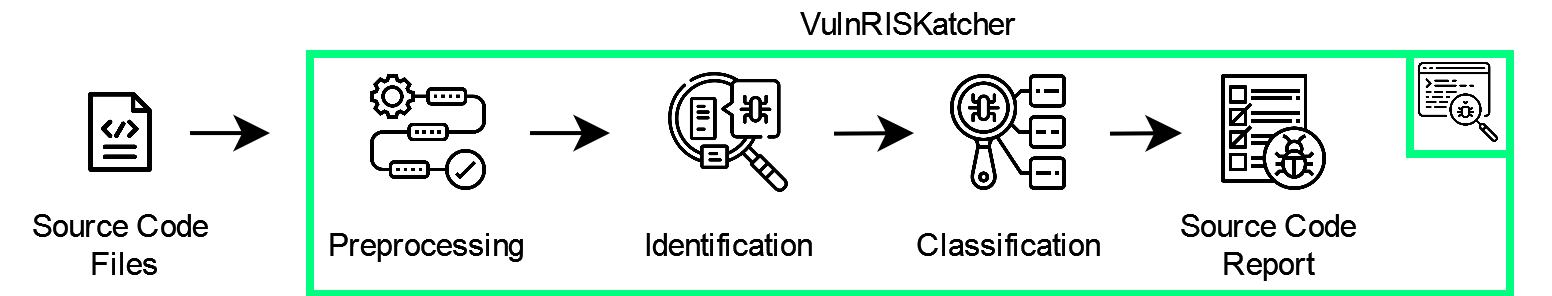}
    \caption{VulnRISKatcher proposed pipeline.}
    \label{fig:VRK_pipeline}
\end{figure}

Fig. \ref{fig:VRK_pipeline} illustrates %the conceptual
pipeline of VulnRISKatcher. 
The user provides either all or part of the code, along with the programming language for accurate data processing. Following this, the tool will preprocess the input data by cleaning the code and converting it into discrete units. The preprocessing stage entails breaking down the code provided by the user into smaller units that can be analyzed. 
The preprocessed code is then analyzed to identify patterns and characteristics associated with vulnerabilities. Then, a classification of the potential vulnerabilities in the provided code is performed. Finally, a source-code report identifying and classifying the code vulnerabilities is presented to the user.

\subsection{CodeAssert}

The most commonly utilized AST tools attempt to automate the software testing process via scripting, as described in Section~\ref{sec:SoTA}. These scripts define multiple tests which attempt to validate a software at multiple levels. However, they still require the manual effort of analyzing the code to create test cases, test data, and the expected results based on the tester's understanding of the SUT.

CodeAssert builds on top of the traditional AST process model by adopting the tools that already use scripting as their AST method. However, it further expands this model by automating the generation of the test scripts, which include all information that is traditionally provided by testers except for the expected result. Naturally, this is a white-box software testing tool that leverages Natural Language Processing (NLP) to analyze source code files and extract all possible test cases. It  achieves 100\% code coverage in an automated manner, ensuring the quality of the System Under Test (SUT) at both the unit and integration level.

\begin{figure}
    \centering
    \includegraphics{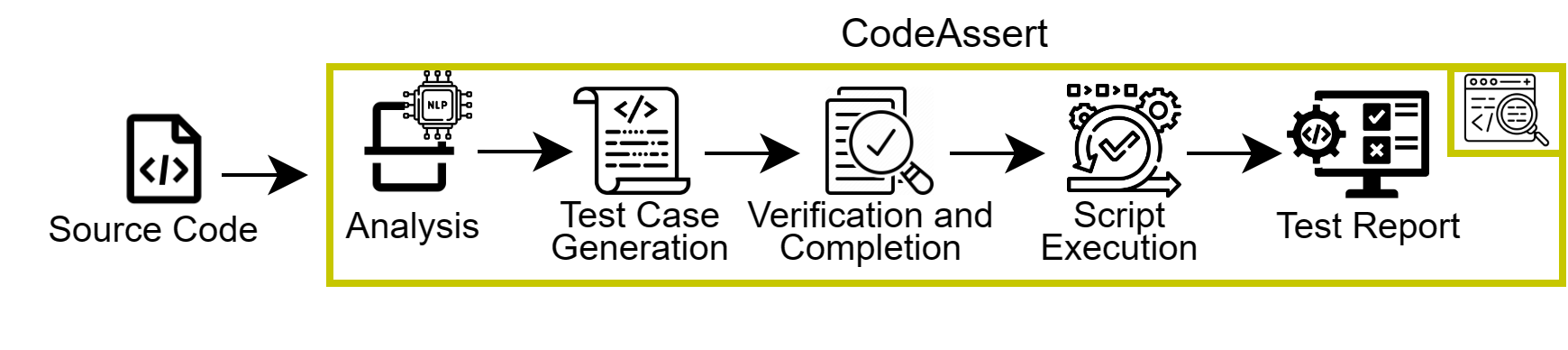}
    \caption{CodeAssert pipeline.}
    \label{fig:CA_pipeline}
\end{figure}

Fig.~\ref{fig:CA_pipeline} describes a simplified execution pipeline of CodeAssert. The user provides the testing tool access to the source code files and selects the type of testing (i.e. unit or integration) that should be conducted. Since it analyzes the source code files, the user must specify the programming language of the source code and the respective scripting tool, to generate the scripts in the correct format. CodeAssert then proceeds to use NLP to analyze the source code and identify all conditional statements. Then the test cases are written based on the identified conditions and the input values are generated accordingly. Afterwards, 
the generated testing scripts are reviewed by the tester, who proceeds to populate the expected value for each identified test case. Finally, the test cases are computed with the chosen scripting AST tool and a testing report is presented to the tester.

\section{Conclusions}
\label{sec:conclusion}

Considering the impact that software has in the people's daily lives, it is essential to ensure its quality in order to avoid disastrous events caused by faulty or vulnerable software. This work presents TestLab, an intelligent AST framework. Its purpose is  to improve software testing in multiple testing levels from different testing perspectives relying on AI-driven methods to achieve automation. 
TestLab was designed to integrate with the software development cycle and enable continuous, comprehensive testing of the SUT, with the objective of ensuring high-quality output throughout the entire process.

Automated software testing has demonstrated its efficiency in saving time, resources, and costs by automating the most monotonous tasks, enhancing software testing, and elevating the overall quality of software. The authors strongly believe that the AI-driven proposed framework has the potential to further improve testing efficiency and effectiveness, reducing the risk of human error, and enabling a more comprehensive testing coverage.

The framework's future work includes the implementation and experimentation of all its modules to assess their performance and effectiveness, as well as to explore additional testing methods that can benefit from AI automation.

\subsubsection*{Acknowledgements.} The present work was partially supported by the Norte Portugal Regional Operational Programme (NORTE 2020), under the PORTUGAL 2020 Partnership Agreement, through the European Regional Development Fund (ERDF), within project ”Cybers SeC IP” (NORTE-01-0145-FEDER-000044). This work has also received funding from UIDB/00760/2020.

\bibliographystyle{unsrt}
\bibliography{bibliography}

\end{document}